\documentclass[ajp,a4paper,superscriptadress,showkeys,showpacs,
nofootinbib,amsmath,amssymb,secnumarabic,twocolumn]{revtex4}
\usepackage{graphicx}% Include figure files
\usepackage{dcolumn}% Align table columns on decimal point
\usepackage{bm}% bold math
\usepackage{times}

\begin{document}
\preprint{***}
\title{Atomic scale friction between clean graphite surfaces}
\author{Katuyoshi Matsushita}
\affiliation{Institute for Solid State Physics, University of Tokyo,\\ 5-1-5 Kashiwanoha, Kashiwa, Chiba, 277-8581, Japan}
\email{kmatsu@issp.u-tokyo.ac.jp}
\author{Hiroshi Matsukawa}
\affiliation{Department of Physics, College of Science and Engineering, Aoyama Gakuin University,  \\5-10-1 Huchinobe, Sagamihara, Kanagawa, 229-8558, Japan}
\email{hm@phys.aoyama.ac.jp}
\author{Naruo Sasaki}
\affiliation{Department of Applied Physics, Faculty of Engineering,
Seikei University,\\
3-3-1 Kichijoji-kitamachi, Musashino-shi, Tokyo 180-8633, Japan}
\affiliation{Precursory Research for Embryonic Science and Technology (PRESTO),
Japan Science and Technology Corporation (JST)\\
4-1-8 Honcho , Kawaguchi-shi, Saitama 332-0012, Japan}
\email{naru@cello.mm.t.u-tokyo.ac.jp}
\begin{abstract}
We investigate atomic scale friction between clean graphite surfaces by using  molecular dynamics. 
The simulation reproduces atomic scale stick-slip motion and low frictional coefficient, both of which are observed in experiments using frictional force microscope. 
It is made clear that the microscopic origin of low frictional coefficients of graphite lies on the honeycomb structure in each layer, not only on the weak interlayer interaction as believed so far.
%The microscopic mechanism of the low frictional coefficient of graphite is made clear for the first time.
\end{abstract}
\maketitle
Friction is one of the most familiar physical phenomena and has been investigated from long ago due to its importance in various kinds of machinery and for the understanding of dynamics of many  systems in science.
Last two decades, atomic scale friction has been attracted much attention for the  understanding of fundamental mechanisms of macroscopic friction and for many fields of high precision engineering such as nanomachine\citep{Persson}. 
Frictional force microscope (FFM) has played an important role in study of atomic scale friction\cite{Persson,Mate}. Behavior of frictional force in FFM experiments has been studied theoretically and numerically based on the model which consists of a single atom tip and
the potential by a substrate
surface\cite{Fujisawa1,Morita,Sasaki1,Sasaki2,Hols}.
The magnitude of load in FFM experiments in layered materials, however, is much higher than that estimated numerically and theoretically\cite{Mate,Sasaki3}. 
Some groups pointed out that a flake cleaved from the
substrate plays a crucial role in FFM
experiments in such systems\cite{Mate,Morita,Miura1,Miura2,Miura3}. 
It is not appropriate to discuss frictional phenomena with a flake by using the single atom tip model. 
We investigate in this letter atomic scale friction between clean graphite surfaces by molecular dynamics (MD). 
The model employed here simulates friction in the FFM experiments on a highly oriented pyrolytic graphite substrate with a flake. 
The present work enables us to understand results of the FFM experiments more deeply.
Graphite is one of the most important solid lubricants.
It turns out that the microscopic origin of  lubrication property of graphite lies on the honeycomb structure in each layer, not only on the weak interlayer interaction.

\begin{figure}[t]
\begin{center}
\includegraphics[width=\linewidth]{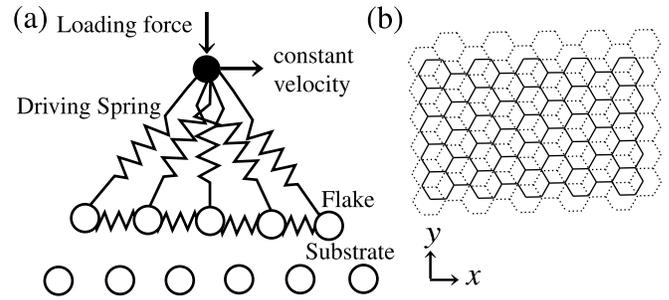}
\end{center}
\caption{(a)The schematic illustration of the model and (b) the flake and  the substrate in the present study. The solid and dashed lines in
 (b) denote bonds in the flake and those in the substrate, respectively. 
\vspace{-0.5cm}}\label{sim2}
\end{figure}

The model consists of a monolayer graphite substrate, a monolayer graphite flake and a spring which drives the flake as shown in fig.~\ref{sim2}(a).
Atoms in the flake obey the following eqs. of motion,
\begin{eqnarray}
m_{\rm C}\ddot{x}_{i}^{\alpha} = - \gamma \dot{x}_{i}^{\alpha}
 - \frac{\partial V_{\rm Sub}}{\partial x_{i}^{\alpha}}
- \frac{\partial V_{\rm I}}{\partial x_{i}^{\alpha}}  
-\frac{k_{\alpha}}{N_{\rm f}}
\left(x_{\rm C}^\alpha-x_{\rm B}^\alpha\right). \label{eqf}
\end{eqnarray}
Here $m_{\rm C}$ is the mass of the carbon atom, 
$x_{i}^{\alpha}$, $\alpha$ $=$ $x$, $y$, $z$, are components of the $i$-th atom position vector and a dot above $x_{i}^{\alpha}$ denotes the time derivative.
The $x$ and $y$-directions are shown in fig.~\ref{sim2}(b) and the $z$-direction is normal to the surface plane of the substrate.
Atoms in the substrate are fixed because their deformations are negligibly
small in the present parameter range.
The first term in the r. h. s. of eq.~(\ref{eqf}) is a damping term proportional to the velocity, which reproduces the energy dissipation to the substrate. 
 We set a damping constant $\gamma$ $=$ 1.14 $\times$ 10$^{-4}$ nNs/m. This value of $\gamma$ holds the stability of the flake for the time step of MD simulation $\Delta t$ = 2.74 $\times 10^{-15}$ sec.
 The second term denotes the force coming from interatomic interactions between the flake and the substrate. $V_{\rm Sub}$ denotes the sum of pair atomic interaction potential between an atom in the flake and that in the substrate.
 We employ the Lennard-Jones $\left(12,6\right)$ potential with a cut off length $R$, outside  of which the interatomic  potential is negligible.
We set Lennard-Jones parameters and the cut-off length as $\varepsilon$ $=$ 0.965 $\times$ 10$^{-2}$ eV, $\sigma$ $=$ 0.340 nm and $R$ $=$ 2.0 nm.
The above values of $\varepsilon$ and $\sigma$ are often employed in the interlayer atomic potential in bulk graphite \cite{Taber}.
The third term denotes the force by the intraflake atomic interactions.
We employ linear spring potentials for the bond length and the angle between bonds of nearest neighbor carbon atoms  and $z$-displacement of an atom against nearest atoms.
They  are employed in the study of lattice vibrations and specific heat of graphite \citep{Yoshimori} and FFM images \citep{Sasaki1,Sasaki2}.
The fourth term denotes the force by the driving spring with the components of spring constants, $k_\alpha$, $\alpha$ = $x$, $y$, $z$. 
The spring corresponds to a tip and a cantilever in FFM. $N_{\rm f}$, $x_{\rm C}^{\alpha}$ and $x_{\rm B}^{\alpha}$ denote the number of carbon atoms in the flake, $\alpha$-components of a center position of the flake mass and a base position of the driving spring, respectively. Here we set $k_x$ = $k_y$ = 1.5 eV/nm$^2$, $k_z$ = 5.0 eV/nm$^2$. Flake size is not known experimentally. 
The sizes of the flake $N_{\rm f}$ in the present calculation are 6, 20, 42 and 110.
The results shown below are those for $N_{\rm f}=110$\cite{size}.
The shape of the flake is shown in fig.~\ref{sim2}(b).
The  size dependence is discussed later.
$\vec{\bm{x}}_{\rm B}$ obeys the following eq. of motion,
\begin{eqnarray}
&&R_{\rm C} m_{\rm C} \ddot{x}_{\rm B}^z = - \gamma \dot{x}_{\rm B}^z
- L + k_{z}\left(x_{\rm C}^z - x_{\rm B}^z \right). \nonumber \\ 
&&\hspace{1.5cm}\dot{x}_{\rm B}^x = V_{\rm S}^x,\,\,\ \dot{x}_{\rm B}^y = V_{\rm S}^y,\label{eqm2}
\end{eqnarray}
Here $R_{\rm C}$, the ratio of the effective mass of the tip and the cantilever to the mass of the carbon atom, is assumed to be 100.0. $L$ and $\left(V_{\rm S}^x,V_{\rm S}^y \right)$ denote the loading force and the lateral driving velocity, respectively. 
The values of these parameters are kept constant during driving. 

Equations of motion,~(\ref{eqf}) and (\ref{eqm2}) are solved numerically by using the Runge-Kutta method. 
In the initial state the system is stable  with the applied loading force $L$ and no lateral expansions of the driving spring. 
The configuration corresponds to the AB stacking structure of bulk graphite as shown in fig.~\ref{sim2}(b).
The frictional force is defined as a driving direction component of the force of the driving spring, $k_{\alpha}\left(x_{\rm B}^{\alpha}-x_{\rm C}^{\alpha}\right)$.

\begin{figure}[t]
\begin{center}
\includegraphics[width=\linewidth]{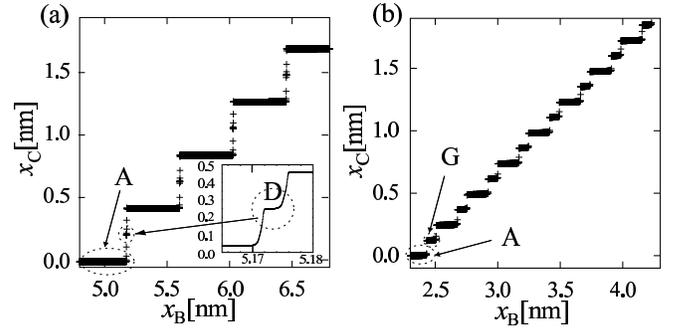}
\end{center}
\caption{$x_{\rm C}$ as a function of $x_{\rm B}$ driven along the (a) $x$  and (b) $y$-directions  for $L$ $=$ 100 nN, $V_{\rm S}$ $=$ 0.22 m/s.  The inset in (a) magnifies the behavior of sticking position D.  The sticking positions A, D and G correspond to those in fig.~\ref{fig3}, respectively. }
\label{ocot}
\end{figure}

It is to be noted that deformations of the flake are negligibly small in the present calculation.
This is because that the interlayer atomic interactions between the flake and the substrate under load are much weaker than the intralayer atomic interactions in the flake.
It is also to be noted that the rotation of the flake around  $\vec{\bm{x}}_{\rm C}$ in the $x$-$y$ plane is inhibited due to the potential barrier against it. 
Due to these two features the dynamics of the center of the flake mass, $\vec{\bm{x}}_{\rm C}$, governs the dynamics of the system.
At first we focus on the typical motions of $\vec{\bm{x}}_{\rm C}$.
Figure~\ref{ocot}
shows the driving direction component of the center of the flake mass, $x_{\rm C}$, as a function of that of the base position of the driving spring, $x_{\rm B}$, for $L$ $=$ 100 nN, $V_{\rm S}$ $=$ 0.22 m/s in the case driven along the $x$ (a) and $y$ (b) directions, respectively. 
The stick-slip motions of $\vec{\bm{x}}_{\rm C}$ are observed. 
These periods agree with the periods of the graphite lattice in each direction, 0.426 and 0.246 nm, respectively.

\begin{figure}[t]
\begin{center}
\includegraphics[width=0.9\linewidth,bb=150 350 410 547]{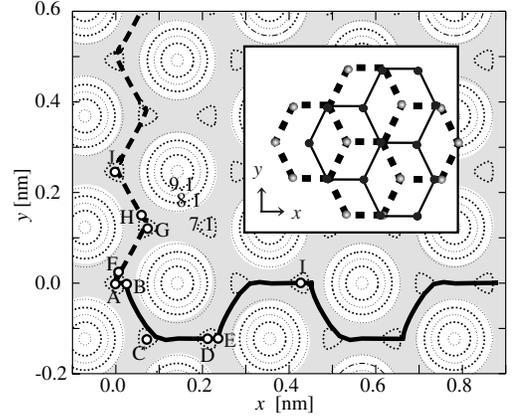}
\end{center}
\caption{The trajectories of $\vec{\bm{x}}_{\rm C}$ for $L$ $=$ 100 nN, $V_{\rm S}$ $=$ 0.22 m/s.
The solid and dashed lines indicate the trajectories of 
$\vec{\bm{x}}_{\rm C}$ driven along the $x$ and $y$-directions, respectively.
The heavy and thin dotted lines indicate the equipotential lines of the substrate potential as a function of
 $\vec{\bm{x}}_{\rm C}$ under the condition that the separation between the flake and the substrate is set to the constant value, 0.273 nm, which is equal to the time averaged value of the separation during driving for $L$ = 100 nN.
The unit of numbers on heavy dotted equipotential lines is eV.
The inset shows the configuration of the flake and the substrate which makes the AB stacking structure of bulk graphite.  
The solid and dashed lines in the inset denote the bonds in the flake and the substrate, respectively.
The shaded region denotes potential valleys.\vspace{-0.5cm}}\label{fig3}
\end{figure}

Figure~\ref{fig3} shows trajectories of $\vec{\bm{x}}_{\rm C}$ in the $x$-$y$ plane for $L$ $=$ 100 nN, $V_{\rm S}$ $=$ 0.22 m/s.
The contour lines denote equipotential lines of the substrate potential.
The motion of $\vec{\bm{x}}_{\rm C}$ starts from the initial position A. 
The solid line in the figure denotes the trajectory of $\vec{\bm{x}}_{\rm C}$ driven along the $x$-direction.
By applying driving force $\vec{\bm{x}}_{\rm C}$ goes slowly to the position B, which is the point of inflection of the total potential.
As soon as $\vec{\bm{x}}_{\rm C}$ arrives at B, $\vec{\bm{x}}_{\rm C}$ slips to the next sticking position close to D through the position C, both of which are the positions of the substrate potential minimum. 
The sticking time around D is much shorter than that around A due to large elongation of the driving spring along the $y$-direction and much longer than the slipping time. The sticking times around A and D are about 99.88 and 0.08 \% of the total period of the stick-slip motion, respectively.
As soon as $\vec{\bm{x}}_{\rm C}$ arrives at E, which is also the point of inflection, $\vec{\bm{x}}_{\rm C}$ slips again close to I, which is equivalent to A. 
Then $\vec{\bm{x}}_{\rm C}$ repeats the periodic stick-slip motion.
%The sticking states around A and D correspond to the regions A and D in fig.~\ref{ocot}(a), respectively.
The dashed line shown in the figure denotes the trajectory of $\vec{\bm{x}}_{\rm C}$ in the case driven along the $y$-direction. 
By applying driving force $\vec{\bm{x}}_{\rm C}$ goes slowly to the position F, which is the point of inflection of the total potential.
As soon as $\vec{\bm{x}}_{\rm C}$ arrives at F, $\vec{\bm{x}}_{\rm C}$ slips to the next sticking position close to G, which is the position of the total  potential minimum. 
The sticking time around G is shorter than that around A due to elongation of the driving spring along the $x$-direction.  
The sticking times around A and G are about 70.0 and 30.0 \% of the period of the stick-slip motion, respectively.
%The sticking times are much longer than the slipping time.
%%
As soon as $\vec{\bm{x}}_{\rm C}$ arrives at H, the flake slips again to the next sticking position close to J, which is equivalent to A.
Then $\vec{\bm{x}}_{\rm C}$ repeats the periodic stick-slip motion and the trajectory of $\vec{\bm{x}}_{\rm C}$ becomes zigzag.
%The sticking states around A and G correspond to the regions A and G in fig.~\ref{ocot}(b), respectively.
%% 
The slipping times are of order of ten pico-seconds.
The slips of $\vec{\bm{x}}_{\rm C}$ occur between positions close to the substrate potential minimum, in which the flake makes the AB stacking structure of the graphite,
%and at the sticking positions with largest sticking time the flake makes the AB stacking structure of bulk graphite with the substrate 
 as shown in the inset of fig.~\ref{fig3}. 
The atomic scale stick-slip motion results from the binding of the flake close to the AB stacking positions of the graphite flake on the graphite substrate. 

\begin{figure}[b]
\begin{center}
\includegraphics[width=1.0\linewidth]{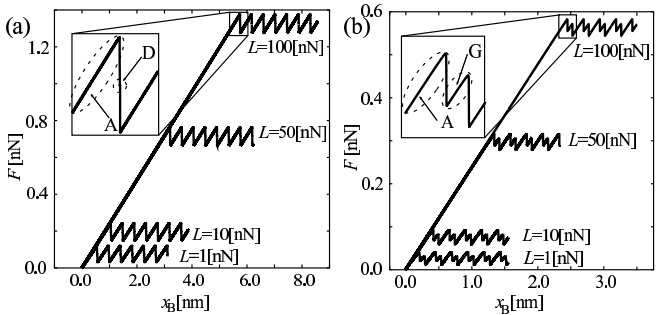}
\end{center}
\caption{The frictional force $F$ as a function of $x_{\rm B}$ for $V_{\rm S}$ $=$ 0.22 m/s and $L$ $=$ 1, 10, 50 and 100 nN and   in the case driven along the (a) $x$ and (b) $y$-directions.}
\label{fig4}
\end{figure}

Figure~\ref{fig4} shows the frictional force $F$ as a function of $x_{\rm B}$ for $V_{\rm S}$ $=$ 0.22 m/s driven along the $x$ (a) and $y$ (b) directions, respectively.
 $F$ initially increases linearly with $x_{\rm B}$ due to the elastic deformation of the driving spring. 
After $F$ takes the maximum value 
the curves exhibit  periodic saw-tooth shapes, which are observed  in  experiments \cite{Mate, Morita, Hoshi, Irr}, due to the periodic stick-slip motions of the flake. 
The regions A, D and G as shown in the insets in fig.~\ref{fig4} (a,b) correspond to sticking states around those in fig.~\ref{fig3}, respectively.

\begin{figure}[b]
\begin{center}
\includegraphics[width=0.9\linewidth]{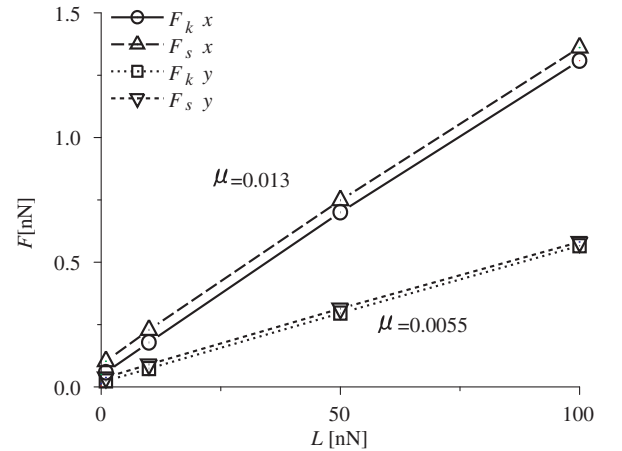}
\end{center}
\caption{$F_k$ and $F_s$, as a function of the loading force. $x$ and $y$ indicate the direction of the driving.} \label{fig5}
\end{figure}

We define the maximum static frictional force $F_s$ and the kinetic frictional force $F_k$ as the maximum frictional force during  driving and a time averaged value of the frictional force during one cycle of the stick-slip motion, respectively.
Figure~\ref{fig5} shows $F_s$ and $F_k$ as a function of the loading force, $L$, driven along the $x$ and $y$-directions.
The dependence of $F_s$ and $F_k$ on $V_{\rm S}$ is weak.
This is because that even in the present calculation the time scale of the driving is much longer than that of atomic one.
The values of the frictional force in fig.~\ref{fig5} are obtained by extrapolating data for $V_{\rm S}$ $=$ 0.22, 0.44 and 0.88 m/s to $V_{\rm S}$ $=$ 0 m/s for the comparison with experiments, in which the typical magnitude of $V_{\rm S}$ is about 40 nm/s.  
$F_s$ and $F_k$ linearly depend on the loading force and have a finite adhesive term $a$, which is the frictional force at $L = 0$ nN.
The frictional coefficient $\mu$ is defined as,
\begin{eqnarray}
\tilde{F} = \mu L + a. \label{deff}
\end{eqnarray}
Here $\tilde{F}$ denotes $F_s$ or $F_k$.
The magnitude of the frictional force in the range of the loading force in fig.~\ref{fig5}
 is very low with the frictional coefficients $\mu_s$ = $\mu_k$ $=$ 0.013 driven along the $x$-direction and $\mu_s$ $=$ $\mu_k$ $=$ 0.0055 driven along the $y$-directions.
Here $\mu_s$ and $\mu_k$ denote the frictional coefficients of $F_s$ and $F_k$, respectively. These values are close to those observed in the experiment, $\mu$ = 0.012 \cite{Mate}.

\begin{figure}[t]
\begin{center}
\includegraphics[width=0.5\linewidth]{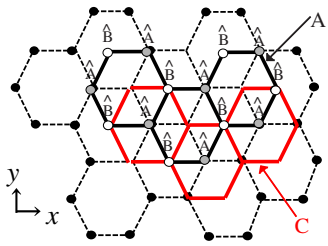}
\end{center}
\caption{The stacking structure of a flake on a substrate. 
The solid and dashed lines denote  bonds in the flake and those in the substrate.
The sites of carbon atoms ${\rm \hat{A}}$ and ${\rm \hat{B}}$ catch different forces of the interatomic interactions between the flake and the substrate.
The configurations A and C correspond with the stacking structures at the positions A and C in fig.~\ref{fig3}.}\label{AB}
\end{figure}

\begin{figure}[t]
\begin{center}
\includegraphics[width=0.7\linewidth]{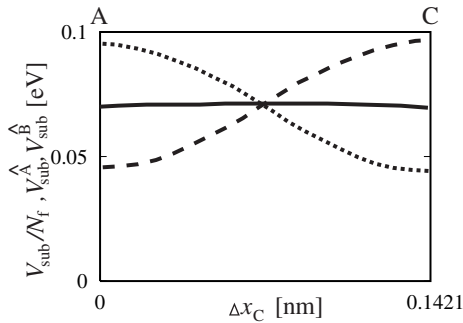}
\end{center}
\caption{The substrate potential during the motion of the flake A $\rightarrow$ C shown in fig.~\ref{fig3}. The solid, dashed and dotted lines denote the potential value  per atom for the flake as a function of a moving-direction displacement of $\vec{\bm{x}}_{\rm C}$, $\Delta x_{\rm C}$, and the potential value as a function of $\Delta x_{\rm C}$ for a single atom at the sites ${\rm \hat{A}}$ and ${\rm \hat{B}}$, respectively.}\label{p}
\end{figure}

Here we discuss the mechanism of the low frictional coefficients of graphite.
It has been believed that the low frictional coefficient of macroscopic graphite surfaces or graphite solid lubricants results from the weak interlayer interaction. 
We obtain, however, the frictional coefficient $\mu$ = 0.099 for a single carbon atom tip on the graphite substrate for the same pressure value with the case of the flake, which is much larger than those in our simulation of the flake and that in the experiment\cite{Mate}. 
The difference of the frictional coefficients indicates an another mechanism of the low frictional coefficients between graphite surfaces.
In the flake there are two kinds of lattice sites ${\rm \hat{A}}$ and ${\rm \hat{B}}$ catching different forces from the substrate as shown in fig.~\ref{AB}. 
The substrate potential for the flake $V_{\rm sub}$ is the sum of the substrate potential for a single atom at the sites ${\rm \hat{A}}$, $V^{\rm \hat{A}}_{\rm sub}$, and that \nopagebreak at the sites ${\rm \hat{B}}$, $V^{\rm \hat{B}}_{\rm sub}$. 
That is, $V_{\rm sub}$ $=$ $N_{\rm \hat{A}}V^{\rm \hat{A}}_{\rm sub}$ $+$ $N_{\rm \hat{B}}V^{\rm \hat{B}}_{\rm sub}$,
where $N_{\rm \hat{A}}$ and $N_{\rm \hat{B}}$ denote the number of the sites ${\rm \hat{A}}$ and ${\rm \hat{B}}$. 
In the flake $N_{\rm \hat{A}}$ = $N_{\rm \hat{B}}$.
Here we consider the case that $\vec{\bm{x}}_{\rm C}$ moves straight between the nearest substrate potential minimums A and C shown in fig.~\ref{fig3}. 
The atoms at sites ${\rm \hat{A}}$ move from the potential minimum to the maximum and those at sites ${\rm \hat{B}}$ move from the maximum to the minimum as shown in fig.~\ref{p}.
During the motion the increase of $V_{\rm sub}^{\rm \hat{A}}$ and the decrease of $V_{\rm sub}^{\rm \hat{B}}$ cancel out each other. 
Then the substrate potential for the flake $V_{\rm sub}$ does not vary in the route between the nearest potential minimums. 
As a result the flat potential valleys appears for the flake in the shaded region shown in fig.~\ref{fig3}. 
The valleys enable the flake to move easily and yield low frictional force.
Due to this mechanism graphite becomes a good lubricant with low frictional coefficient. 
On the other hand the single carbon atom above the graphite substrate has large frictional coefficient because of the absence of such cancellation mechanism of the substrate potential.

The frictional coefficient mainly depends on the trajectory of $\vec{\bm{x}}_{\rm C}$. 
When the flake is  large enough that the interaction between the flake and the substrate is so strong as to overcome the stiffness of the driving spring, the dominant contribution to the acceleration of the flake is flake size independent value, $(\partial_{\rm A} V^{\rm \hat{A}}_{\rm sub}+\partial_{\rm \hat{B}} V^{\rm \hat{B}}_{\rm sub})/2m_{\rm C}$, for the same separation between the flake and the substrate. 
Here $\partial_{\rm \hat{A}, \hat{B}}$ denote the coordinate derivatives at the sites $\rm \hat{A}$ and $\rm \hat{B}$, respectively. 
Then because
the trajectory is independent of the flake size, the frictional coefficient is also independent of the flake size. 
The flake in this simulation is sufficiently large, so that the frictional coefficient is independent of the flake size. 
In fact the difference between the frictional coefficients for $N_{\rm f}$ = 42 and 110 is much small as about 2 \%.

% The sticking time, which is much longer than the slipping time, is observed in FFM experiment for high scan velocity \cite{Hoshi}. Such a long slipping time is not observed in this simulation due to zero temperature and softness of the cantilever and the tip.

In the present work we have investigated atomic scale friction between clean graphite surfaces by numerical simulation. The simulation reproduces atomic scale stick-slip motion and low frictional coefficients of graphite. 
The former is due to the binding of the flake close to the stacking configurations on the substrate. 
The latter results from the cancellation of the forces coming from the substrate between two kinds of lattice sites in the flake, which is discovered  in this letter for the first time within our knowledge.

We express sincere thanks to Profs. K.~Miura and S.~Morita for valuable discussions. The present work is financially supported by Grant-in-Aid for Scientific Research from Japan Society for the Promotion of Science.


\begin{thebibliography}{9}
\bibitem{size} The magnitudes of the load in the FFM experiment  is about 100 times larger than those in the single atom chip simulation for the same frictional force patterns \cite{Sasaki3} .  Hence we employ mainly the flake with the size 110 atoms in the simulation.  In the experimen by Miura et.al the size of the flake is the order of mm\cite{Miura3},  In the latter case the deformation of the flake may be important.
\bibitem{Irr} In the present calculation irregular stick-slip motion is not observed, which is sometimes appear in experiments\cite{Mate,Morita,Hoshi}.
\end{thebibliography}
\end{document}